\documentclass[11pt]{article}
\usepackage[dvips]{graphicx}
\usepackage[utf8]{inputenc}
\usepackage[english]{babel}
\usepackage[T1]{fontenc}

\usepackage[title,titletoc]{appendix}
\usepackage{graphicx}
\usepackage{color}
\usepackage{float}
\usepackage{tabularx}
\usepackage{indentfirst}
\usepackage{hyperref}
\usepackage{pdfpages}
\usepackage{fancybox}
\usepackage{setspace}

\usepackage[T1]{fontenc}

\usepackage[table,dvipsnames,xcdraw]{xcolor}  
\usepackage{tabularx}
\usepackage{booktabs}
\usepackage{enumitem}

\usepackage{pdflscape}  
\usepackage{afterpage} 
\usepackage{float}  
\usepackage{placeins} 

\usepackage{graphicx}
\usepackage{pgfgantt}
\usepackage{xcolor}
\usepackage[normalem]{ulem} 
\usepackage{color,soul} 

\usepackage{multicol}
\usepackage{graphicx}

\usepackage{subfigure}


\newenvironment{capalayout}{
    \setlength{\topmargin}{-3.0cm}
    \setlength{\footskip}{0cm}
    \thispagestyle{empty}
} 

\newcommand{\ISSNno}{0103-9741}
\newcommand{\MCCSeqAno}{03/2021}
\newcommand{\TituloCapa}{On Psychometric Instruments in Software Engineering Research: An Ongoing Study}

\newcommand{\AutorANome}{Danilo Almeida Felipe}
\newcommand{\AutorAemail}{dfelipe@inf.puc-rio.br}

\newcommand{\AutorCNome}{Marcos Kalinowski}
\newcommand{\AutorCemail}{kalinowski@inf.puc-rio.br}

\begin{document}

\begin{capalayout}
        \centering
        \includegraphics[keepaspectratio,width=14.7cm]{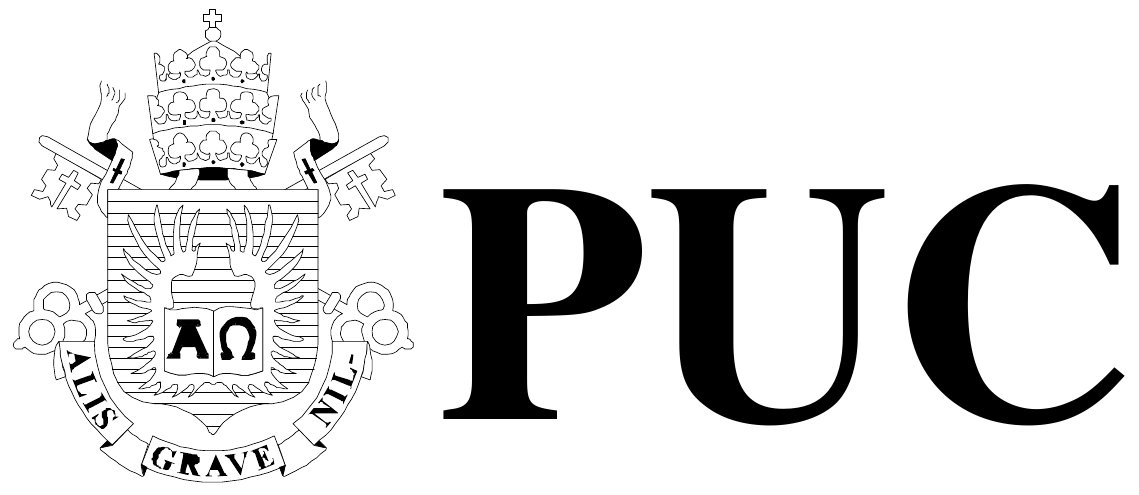}
        
        \medskip
    
        \setlength\fboxsep{1pt}
        \shadowbox{\fbox{\begin{minipage}[h]{14.5cm}
            \begin{center}
                \doublespacing
                \fontfamily{phv} \fontsize{14}{16} \selectfont
                \medskip
                ISSN \ISSNno
                
                \bigskip
                Monografias em Ciência da Computação 
                
                n\textordmasculine \, \MCCSeqAno
                
                \bigskip
                \medskip
                \fontsize{18}{20}\selectfont
                \textbf{\TituloCapa}
                
                \bigskip
                \fontsize{14}{15}\selectfont
                \textbf{\AutorANome} \\
                \textbf{\AutorCNome}
                
                \bigskip
                
                \medskip
                Departamento de Informática
                \bigskip
            \end{center}
        \end{minipage}}}
        
        \bigskip
        \bigskip
        
        \begin{minipage}[h]{14.5cm}
            \doublespacing
            \fontfamily{phv}
            \begin{center} 
                \fontsize{12}{14}\selectfont
                \textbf{PONTIFÍCIA UNIVERSIDADE CATÓLICA DO RIO DE JANEIRO} \\
                \textbf{RUA MARQUÊS DE SÃO VICENTE, 225 - CEP 22451-900} \\
                \textbf{RIO DE JANEIRO - BRASIL}
            \end{center}
        \end{minipage}
        \newpage
\end{capalayout}
\thispagestyle{empty}

\begin{flushleft}
    \begin{tabular}{p{11.1cm}r}
        Monografias em Ciência da Computação, No. \MCCSeqAno & ISSN: \ISSNno \\
        Editor: Prof. Carlos José Pereira de Lucena & Agosto, 2021
    \end{tabular}
\end{flushleft}
\LARGE
\bigskip

\begin{center}
    {\bf \TituloCapa}
\end{center}
\normalsize
\bigskip

\begin{center}
    {\bf \AutorANome, \AutorCNome}
\end{center}

\begin{center}
   \AutorAemail, \AutorCemail
\end{center}


\noindent {\bf Abstract.} [Context] Although software development is an inherently human activity, research in software engineering (SE) has long focused mostly on processes and tools, failing to recall about the human factors behind. Even when explored, researchers typically do not properly use psychology background to better understand human factors in SE, such as the psychometric instruments, which aim to measure human factors.
[Objective] Our goal is to provide a critical review on the use of psychometric instruments in SE research regarding personality.
[Method] We present a two-step study. First, a systematic mapping of the literature in order to generate a catalog of the psychometric instruments used; second, a preliminary survey to be conducted with social sciences researchers to assess their adoption in SE research.
[Results and Conclusion] The results so far are quite initial. The next steps direct us to finish the data extraction to finalize the catalog (systematic mapping) and to refine the survey design and apply it with social sciences researchers.
\medskip
\medskip


\bigskip
\noindent {\bf Resumo.} [Contexto] Embora o desenvolvimento de software seja uma atividade humana, a pesquisa em Engenharia de Software (ES) tem concentrado esforços majoritariamente em processos e ferramentas, esquecendo-se dos fatores humanos por trás. Ainda quando explorados, os pesquisadores não tem adotado adequadamente referencial da psicologia para entender melhor os fatores humanos em ES, bem como dos instrumentos psicométricos, que visam medir algum tipo de fator humano.
[Objetivo] Nosso objetivo é fornecer uma revisão crítica sobre o uso de instrumentos psicométricos nas pesquisas em ES em relação à personalidade.
[Método] Apresentamos um estudo em duas etapas. Primeiro, um mapeamento sistemático da literatura para gerar um catálogo dos instrumentos psicométricos utilizados; segundo, um survey preliminar a ser realizado com pesquisadores de ciências sociais/humanas para avaliar sua adoção na pesquisa em ES.
[Resultados e Conclusões] Os resultados até agora são iniciais. Os próximos passos nos direcionam a finalizar a extração de dados para finalizar o catálogo (mapeamento sistemático) e refinar o survey para conduzí-lo com pesquisadores de ciências sociais/humanas.
\medskip
\medskip

\noindent {\bf Palavras-chave:} engenharia de software comportamental; personalidade; mapeamento da literatura; survey \\
\newpage
\pagenumbering{roman} \setcounter{page}{2}
\vspace*{\fill}
\begin{flushleft}
    \textbf{In charge of publications:} \\
    PUC-Rio Departamento de Informática - Publicações \\
    Rua Marquês de São Vicente, 225 - Gávea \\
    22453-900 Rio de Janeiro RJ Brasil \\
    Tel. +55 21 3527-1516 Fax: +55 21 3527-1530 \\
    E-mail: publicar@inf.puc-rio.br \\
    Web site: http://bib-di.inf.puc-rio.br/techreports/ \\
    \end{flushleft}

\newpage

\renewcommand{\contentsname}{Table of Contents}
\renewcommand{\appendixname}{Annex}
\tableofcontents
\newpage

\newcommand{\eg}{\textit{e.g.}, }
\newcommand{\ie}{\textit{i.e.}, }
\newcommand{\etal}{\textit{et al.} }

\newcommand{\rqmap}{\textit{What are the psychometric instruments used regarding personality in SE research?}}

\newcommand{\rqsurvey}{\textit{How do social sciences researchers perceive the adoption of psychometric instruments regarding personality in SE research?}}

\newcommand{\rqmapA}{\textit{What are the objectives of the studies?}}

\newcommand{\rqmapB}{\textit{For what purposes have psychometric instruments been used in the studies?}}

\newcommand{\rqmapC}{\textit{Which types of research do the studies refer to?}}

\newcommand{\rqmapD}{\textit{Which types of empirical studies have been conducted?}}

\newcommand{\rqmapE}{\textit{What are the limitations faced by the use of psychometric instruments reported in the studies?}}

\pagenumbering{arabic} \setcounter{page}{1}

\section{Introduction}
\label{sec:introduction}

Software Engineering (SE) activities are primarily performed by humans. However, many empirical studies have only focused on proposing new methods and technologies to support SE activities leaving human and social factors behind them underexplored \cite{feldt_towards_2008}, not allowing a more holistic view of the area.

Behavioral Software Engineering (BSE) is the body of knowledge of SE research that attempts to understand human aspects related to the activities of software engineers, software developers, and other stakeholders \cite{lenberg_behavioral_2015}. The topic has been the subject of recent research in the SE domain. Nevertheless, because it is relatively immature, some approaches adopted misled researchers mainly by not properly combining SE research with social sciences backgrounds, such as psychology, to address human factors \cite{graziotin_affect_2015}. Graziotin \etal\cite{graziotin_what_2018} enlighten a set of existing research on relating developer happiness to productivity, software quality, and social interactions.

Furthermore, in SE research, it is inherent to note that measurement activities are an essential part. In empirical studies the researcher(s) must be sure when adopting ways of measuring the study's resources in question. These resources can be personnel, hardware, or software for an activity or process \cite{wohlin_experimentation_2012}

BSE research has encouraged the use of psychometric instruments as support to the understanding of human factors in a more systematic way \cite{feldt_towards_2008, lenberg_behavioral_2015}. In its turn, Psychoempirical Software Engineering proposed in Graziotin \etal\cite{graziotin_affect_2015} deals with “denoting research in Software Engineering with proper theory and measurements from psychology”. A problem addressed by the authors is the misuse of theoretical backgrounds of psychology, such as assuming a theory as the only truth in the research foundation; and also the improper use of psychometric instruments, some not validated from psychology or used to evaluate wrong human factors. The authors propose guidelines to assist researchers in addressing these problems.

However, information about how psychometric instruments are adopted in SE research remains vague and dispersed in many studies. As far as we know, one study has partially synthesized this knowledge concerning a specific construct, personality\footnote{Personality is one of the most studied concepts in BSE research as pointed by Lenberg \etal\cite{lenberg_behavioral_2015}.}: a period of forty years (1970 - 2010) about personality in SE research is mapped by Cruz \etal\cite{cruz_forty_2015}. Still, there is only a brief discussion and characterization of the instruments and their use in the software engineering context (\ie education and pair programming), missing a critical assessment. This status quo remains unchanged more than ten years later and deserves to be challenged.

There is a need to find out whether SE research over the years has been adopting these instruments coherently with their proposed form of use in the social sciences. Although some authors have outlined this on a smaller scale \cite{mcdonald_who_2007, usman_use_2019}, a large-scale study has not been done to get a big picture.

In order to synthesize this knowledge in a structured manner, \textbf{our objective is to present an overview on the use of psychometric instruments in SE research on personality}\footnote{Hereafter we commonly refer to psychometric instruments related to personality in SE research simply as ``psychometric instrument(s)''.}, with a focus on classifying the \textit{type of research}, \textit{empirical evaluations}, \textit{objective of the studies}, and \textit{purposes of the psychometric instruments}. We intend to consolidate our findings in a catalog through a systematic mapping study (secondary study). Also, to \textbf{assess the use of these instruments from the point of view of social sciences researchers} by conducting a survey (primary study).

More specifically, we address the following Research Questions (RQ):

\begin{itemize}
    \item \textbf{RQ1: }\rqmap
    \item \textbf{RQ2: }\rqsurvey
\end{itemize}

To the best of our knowledge, this study is the first one focused on creating a catalog of psychometric instruments used in SE research and assessing them with social sciences researchers in a larger scale. The remainder of this study manuscript is organized as follows. Section \ref{sec:background} presents the background concepts and related work. Section \ref{sec:smp} presents the systematic mapping protocol and preliminary results. Section \ref{sec:psp} presents our preliminary survey protocol. Finally, section \ref{sec:timetable} presents our concluding remarks.

\section{Background and Related Work}
This section provides background information and related work.
\label{sec:background}

\subsection{Behavioral Software Engineering (BSE)}
\label{sec:background-bse}

BSE is defined as “the study of cognitive, behavioral and social aspects of software engineering performed by individuals, groups or organizations” \cite{lenberg_behavioral_2015}. It involves dealing with existing relationships between SE and disciplines from social sciences, such as \textit{work and organizational psychology}, the \textit{psychology of programming}, and \textit{behavioral economics} to get a broader understanding of SE practices.

Although software is developed by humans, for a long-time, SE research has focused intensively on the technical aspects (such as processes and tools) and less on human and social aspects \cite{feldt_towards_2008}. A summarization conducted in Graziotin \etal\cite{graziotin_consequences_2017} points out existing research, in the scope of BSE, on relating developer happiness to productivity, software quality, and also social interactions.

Still, in the context of BSE, Lenberg \etal\cite{lenberg_behavioral_2015} present a definition for the body of knowledge research described earlier and also conducted a systematic literature review based on the definition. The findings report lack of research in some SE knowledge areas (\ie requirements, design, and maintenance) and rare collaboration between SE and social science researchers.

Cruz \etal\cite{cruz_forty_2015} performed a systematic mapping on personality in SE research. In addition to reporting on the most common SE topics addressed, such as \textit{education} and \textit{extreme programming}, they report which personality tests were most commonly used, which resembles this work. However, the authors only collected and reported brief information on personality-related psychometric instruments without deep discussion about them regarding their use in SE. They provide a valuable list of instruments and relate them to some SE topics, but lack a critical review involving the lens of social scientists.

Further, Usman and Minhas \cite{usman_use_2019} investigate ethical topics raised by McDonald and Edwards \cite{mcdonald_who_2007} on the adoption of MBTI (Myers-Briggs Type Indicator) based tests, a psychometric instrument, in a sample of 8 studies obtained in the final set of Cruz \etal\cite{cruz_forty_2015} published after 2007, and complemented with 7 studies returned in string-based search in Scopus\footnote{https://www.scopus.com/} in the years of 2016 and 2017, totaling a sample of 15 studies. The authors found problems in all of them, including the reliability and validity of MBTI (there are different versions of this psychometric instrument). The authors also highlight possible causes, such as not exploring literature guidelines and lack of collaboration with social science researchers. However, the study reported is initial and limited to analyzing only the use of MBTI in a small sample of studies.

Our study intends to conduct an extensive analysis, with more cataloged psychometric instruments, exploring their relation with SE theory and using expert knowledge to assess these instruments' use. 


\subsection{Psychometrics in Software Engineering}
\label{sec:background-psySE}

According to Michell \cite{michell_measurement_1999}, “psychometrics is concerned with theory and techniques for quantitative measurement in psychology and social sciences” [Michell, 1999 \textit{apud} Feldt et al., 2008], in addition Feldt \etal\cite{feldt_towards_2008} state that “... in practice, this often means the measurement of knowledge, abilities, attitudes, emotions, personality, and motivation”. In the present study, we focused on personality.

The use of psychometric instruments in SE has been encouraged, especially in empirical research, as a way to emphasize hitherto unexplored human factors and to help understand how they affect the research landscape \cite{feldt_towards_2008}.

As far as we know, McDonalds and Edwards \cite{mcdonald_who_2007} is the first study that manifests attention on the use of psychometric instruments in SE research, beyond providing guidelines for the use of two of them (MBTI and 16PF). In addition, one of the authors is from social sciences and a certified professional regarding these instruments.

Still under discussion, Graziotin \etal\cite{graziotin_affect_2015} claim that the use of psychometric instruments should be cautious, recommending that they should be validated; in addition to the proper theoretical background used. The authors then propose the \textit{Psychoempirical Software Engineering} that aims “to denote research in SE with proper theory and measurement from psychology”. In the same study, the authors provide guidelines for the proposal.

More recently, Usman and Minhas \cite{usman_use_2019} show results, despite several limitations, that the use of psychometric instruments in SE is inadequate. The authors emphasize who should apply them, but we consider that trained professionals are the best decision according to the little evidence and guidelines in literature so far.

Although we are researchers without formal qualifications in social sciences, we rely on consolidated methodological tools of SE to conduct this study and plan to involve social science scientists in our research. Given the gathering of evidence collected to answer \textbf{RQ1}, we intend to use the experts' opinion to answer \textbf{RQ2} in order to allow for a critical assessment. The next section presents our systematic mapping protocol and preliminary results.

\section{Systematic Mapping Protocol}
\label{sec:smp}

Systematic mapping is a method to build a classification scheme of an area providing a visual summary of the state of research in a structured way \cite{petersen_systematic_2008}. It aims at providing an auditable and replicable process with minimal bias.

This section describes each step of our research method based on guidelines in the literature. Subsection \ref{sec:smp-rqs} introduces the mapping goal and research questions. Subsection \ref{sec:smp-search} describes the search strategy for collecting new evidence. Subsection \ref{sec:smp-studysel} presents the study selection criteria and discusses quality assessment. The subsection \ref{sec:smp-def} presents the Data Extraction Form and our classification scheme, and subsection \ref{sec:smp-progress} presents preliminary results.

\subsection{Mapping Goal and Research Questions}
\label{sec:smp-rqs}

Our systematic mapping aims at providing an overview of the use of psychometric instruments in SE research. To guide our investigation, and to obtain an overview of the state-of-the-art, trends and gaps, we describe the main Research Question (RQ) as follows:

\textbf{RQ1: }\rqmap

Based on the main RQ, we derived five secondary RQs in order to further characterize the field as follows.

\begin{itemize}
    \item \textbf{RQ1a: }\rqmapA
    \item \textbf{RQ1b: }\rqmapB
    \item \textbf{RQ1c: }\rqmapC
    \item \textbf{RQ1d: }\rqmapD
    \item \textbf{RQ1e: }\rqmapE
\end{itemize}

In the following subsection the search strategy is presented. It was developed by the first author and reviewed by the second author.

\subsection{Search Strategy and Data Source}
\label{sec:smp-search}

\subsubsection{Existing Mapping Study and the Need to Update}

This mapping started being conducted in a standard way of conducting secondary studies (string-based search in digital libraries with snowballing steps), according to consolidated literature \cite{petersen_systematic_2008, petersen_guidelines_2015, kitchenham_guidelines_2007, mourao_investigating_2017}. Until new guidelines emerged, and we noticed that they could help conduct our study \cite{mendes_when_2020, wohlin_guidelines_2020}, given the awareness we had about comprehensive secondary studies on human factors in SE that could be updated \cite{cruz_forty_2015, lenberg_behavioral_2015}.

We defined Cruz \etal\cite{cruz_forty_2015} as a candidate for updating as we wanted to start our immersion into BSE using a narrower scope, focused on personality, to allow a complete overview and a precise critical assessment. Cruz \etal\cite{cruz_forty_2015} identified 90 papers published within a time range of 40 years, more than 10 years ago. Lenberg \etal\cite{lenberg_behavioral_2015} define 55 BSE concepts (\eg \textit{personality}, \textit{job satisfaction}, \textit{communication}, etc.) in a large scale study that considered 250 papers. The narrower focus by Cruz \etal\cite{cruz_forty_2015} would also allow us to apply our updating search strategy (discussed in subsection \ref{sec:smp-update}) with reasonable efforts.

We also argue that personality is a BSE concept presented in \cite{lenberg_behavioral_2015} as one of the most studied together with others (such as \textit{group composition}, \textit{communication}, and \textit{organizational culture}). We believe that updating Cruz \etal\cite{cruz_forty_2015} yields significant results regarding our objectives described in section \ref{sec:introduction}, which is also within BSE's scope. Thus, we decided to update the mapping study by Cruz \etal\cite{cruz_forty_2015}.

\begin{figure}[h]
\caption{Framework recommended by Mendes \etal\cite{mendes_when_2020}, adapted from Garner \etal\cite{garner_when_2016}}
\centering
\includegraphics[width=10cm]{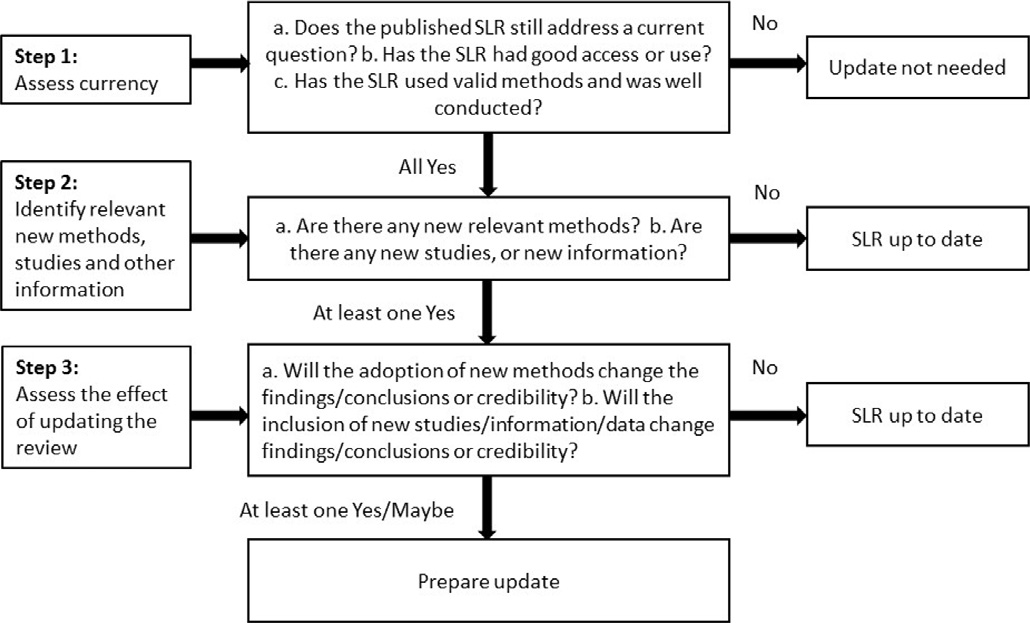}
\label{fig:3pdf-framework}
\end{figure}

To evaluate the need of updating Cruz \etal\cite{cruz_forty_2015}, we used the framework recommended by guidelines to evaluate the possibility of updating secondary studies in SE \cite{mendes_when_2020}\footnote{This study proposes guidelines for updating Systematic Literature Reviews (SLR). Despite this, we believe that our mapping goal is comprehensive enough to apply in the context of systematic mappings.}. We conducted the evaluation process by answering the same seven RQs used in Mendes \etal\cite{mendes_when_2020}, listed and answered hereafter (Steps 1.a. to 3.b.) and ilustrated in Figure \ref{fig:3pdf-framework}.

\textbf{Step 1.a.: }\textit{Does the published SLR still address a current question?}

Yes. Human factors, such as personality, have been the subject of SE research despite being historically poorly explored \cite{graziotin_what_2018}. A list of arguments follows: recent efforts to consolidate a body of knowledge in SE as proposed by Lenberg \etal\cite{lenberg_behavioral_2015} defining BSE; existing conferences such as CHASE (Workshop on Cooperative and Human Aspects of Software Engineering), subsidized by ICSE (International Conference on Software Engineering)\footnote{http://www.icse-conferences.org/}, the largest conference in SE; award-winning papers on the topic in main software engineering related venues, such as \cite{graziotin_what_2018}, awarded in the Journal of Systems and Software\footnote{https://www.journals.elsevier.com/journal-of-systems-and-software}. All these arguments show the importance of human factors as a relevant research topic in SE.

\textbf{Step 1.b.: }\textit{Has the SLR had good access or use?}

Yes. Like Mendes \etal\cite{mendes_when_2020}, we used in the cut-off point the same yearly average citation value of 6.82 documented by Garousi and Fernandes \cite{garousi_highly-cited_2016} to consider a paper for good access or use. In August of 2020, Cruz \etal\cite{cruz_forty_2015} had a yearly average citation value of ~30.8 in Google Scholar.

\textbf{Step 1.c.: }\textit{Has the SLR used valid methods and was it well-conducted?}  

Yes. Regarding the methods, Cruz \etal\cite{cruz_forty_2015} present an extension of preliminary results published in Cruz \etal\cite{cruz_personality_2011} with some improvements, such as a refined search string increasing the sensitivity and coverage; adding backward snowballing steps; review of RQs and extended presentation of results.
Finally, the authors present clear steps in their mapping protocol and are based on well-recognized guidelines for secondary studies in SE \cite{kitchenham_guidelines_2007}.

\textbf{Step 2.a.: }\textit{Are there any new relevant methods?}

Yes. Concerning new methods about our mapping protocol, we adopted guidelines presented as the best way to search for evidence to update secondary studies in SE \cite{mendes_when_2020}. We believe that we used good literature references to help answer our RQs and, consequently, reflect on the results' presentation. However, different from Cruz \etal\cite{cruz_forty_2015} mapping study, our focus is totally on the psychometric instruments within SE research, not a characterization of an aspect (personality) in general.

\textbf{Step 2.b.: }\textit{Are there any new studies or new information?}

Yes. The papers included in the original study had each a considerable number of citations in a preliminary verification in Google Scholar. In addition to having a five-year time interval since the publication of the mapping and the conduction of the update in this present study (2015 to 2020), beyond the period of ten years (2011 to 2020), not incorporated by Cruz \etal\cite{cruz_forty_2015}.

\textbf{Step 3.a.: }\textit{Will the adoption of new methods change the findings, conclusions or credibility?}

Yes, potentially. We adopted a new method concerning the mapping protocol and addressed different RQs to get a big picture of psychometric instruments in SE research, which we believe generates new and important findings. Cruz \etal\cite{cruz_forty_2015} has a relevant RQ on the psychometric instruments/personality tests used, but there is little discussion of its results beyond listing and counting frequencies.

\textbf{Step 3.b.: }\textit{Will the inclusion of new studies/information/data change findings, conclusions or credibility?} 

Yes. Regarding new potential findings, we had prior knowledge of a series of studies \cite{graziotin_affect_2015, graziotin_consequences_2017, graziotin_what_2018} used as control papers. These studies are not covered by Cruz \etal\cite{cruz_forty_2015} because they were published later. They discuss the use of psychometric instruments in SE, and on theoretical basis of other areas (such as social sciences and psychology), which can support SE research in general.

Next, we describe the search strategy to collect new evidence from a secondary study update in SE.

\subsubsection{Strategy to Collect New Evidence}
\label{sec:smp-update}

We adopted the guidelines proposed in Wohlin \etal\cite{wohlin_guidelines_2020} as a strategy to search for new evidence from a secondary study update. They are the following:

\begin{itemize}
    \item \textbf{Use a seed set containing the original secondary study and its included primary studies:} Cruz \etal\cite{cruz_forty_2015} included 90 papers in their final set. However, one of them was excluded (S86) because it is a book chapter, and we did not find evidence of publication in a scientific journal or conference to be approved in IC1 (see Table \ref{IE-table}). As suggested, the secondary study itself was included, obtaining a seed set of 90 studies.
    \item \textbf{Use Google Scholar to search for papers and apply Forward Snowballing (FS), without iteration:} We use the Publish or Perish 7 tool \cite{harzing_publish_2020} to assist this step. The tool has features related to bibliometric analysis, in which one is to retrieve citations from publications using Google Scholar (a FS feature). Thus, we conducted the FS in the seed set using the tool in August 2020 and exported the results for treatment in JabRef, a bibliographic reference manager. Also, a new FS step was conducted in January 2021 to ensure full 2020 indexation. All screening steps were conducted using JabRef.
    \item \textbf{Include more than one researcher in the initial screening to minimize the risk of removing studies that should be included (false negatives):} One researcher was included to assist in the initial screening of studies and discussions were held with a third researcher.
\end{itemize}

\subsection{Study Selection}
\label{sec:smp-studysel}

Petersen \etal\cite{petersen_guidelines_2015} argue that only studies that are relevant to answer the RQs must be considered. The inclusion and exclusion criteria applied to filter the raw set of studies from FS are presented in Table \ref{IE-table}.

\begin{table}[h] \footnotesize
 \centering
 {\renewcommand\arraystretch{1}
 \caption{Inclusion and exclusion criteria.}
 \label{IE-table}
 \begin{tabular}{ l l }
  \cline{1-1}\cline{2-2}  
    \multicolumn{1}{|p{2cm}|}{\textbf{Criteria} \centering } &
    \multicolumn{1}{p{9cm}|}{\textbf{Description} \centering }
  \\  
  \cline{1-1}\cline{2-2}  
    \multicolumn{1}{|p{2cm}|}{IC1} &
    \multicolumn{1}{p{9cm}|}{Papers published in journals and conferences reporting software engineering research using psychometric instruments regarding personality that were published after 2010.}
  \\  
  \cline{1-1}\cline{2-2}  
    \multicolumn{1}{|p{2cm}|}{EC1} &
    \multicolumn{1}{p{9cm}|}{Papers that are not written in English.}
  \\  
  \cline{1-1}\cline{2-2}  
    \multicolumn{1}{|p{2cm}|}{EC2} &
    \multicolumn{1}{p{9cm}|}{Grey literature. Such as books, theses (bachelor's degree, MSc or PhD), technical reports, occasional papers, and manuscripts without peer-review evidence.}
  \\  
  \cline{1-1}\cline{2-2}  
    \multicolumn{1}{|p{2cm}|}{EC3} &
    \multicolumn{1}{p{9cm}|}{Papers that are only available in the form of abstracts, posters, and presentations.}
  \\ 
  \cline{1-1}\cline{2-2}  
    \multicolumn{1}{|p{2cm}|}{EC4} &
    \multicolumn{1}{p{9cm}|}{Papers that did not include in their title or abstract terms defined in Cruz \etal\cite{cruz_forty_2015} regarding personality. There are: ‘‘personality’’, ‘‘psychological typology’’, ‘‘psychological types’’, ‘‘temperament type’’, and ‘‘traits’’.}
  \\  
  \cline{1-1}\cline{2-2}  
    \multicolumn{1}{|p{2cm}|}{EC5} &
    \multicolumn{1}{p{9cm}|}{Studies addressing other psychometric constructs (\eg behavior, cognition, abilities, roles, etc.)}
 \\
 \hline
 \end{tabular} }
\end{table}
\FloatBarrier

The purpose of this mapping study is to provide an overview of the use of psychometric instruments in SE research published in peer-reviewed venues. Therefore, we focus on classifying the type of contribution, the use of psychometric instruments, and the type of research to understand the overall publication landscape, without applying a formal quality assessment. The procedure here involves reading titles and abstracts and looking for evidence of psychometric instruments. If it is not enough for clarification, the paper’s introduction and conclusion will be read. Still, if not sufficient, the full text of the study will be read.

\subsection{Data Extraction and Classification Scheme}
\label{sec:smp-def}

The data extracted from each paper of the final set is shown in Table \ref{DEF}.

\begin{table}[h] \footnotesize
 \centering
 {\renewcommand\arraystretch{1}
 \caption{Data Extraction Form}
 \label{DEF}
 \begin{tabular}{ l l }
  \cline{1-1}\cline{2-2}  
    \multicolumn{1}{|p{3cm}|}{\textbf{Information} \centering } &
    \multicolumn{1}{p{7cm}|}{\textbf{Description} \centering }
  \\  
  \cline{1-1}\cline{2-2}  
    \multicolumn{1}{|p{3cm}|}{Study Metadata} &
    \multicolumn{1}{p{7cm}|}{Paper title, author's information, venue, psychometric instrument (name, version, and application process), and year of publication.}
  \\  
  \cline{1-1}\cline{2-2}  
    \multicolumn{1}{|p{3cm}|}{Objective (RQ1a)} &
    \multicolumn{1}{p{7cm}|}{Study objective: we will use open coding \cite{stol_grounded_2016} to extract information.}
  \\  
  \cline{1-1}\cline{2-2}  
    \multicolumn{1}{|p{3cm}|}{Purpose of the psychometric instrument in the study (RQ1b)} &
    \multicolumn{1}{p{7cm}|}{What constructs represent the purpose of the psychometric instrument in the study. In SE, constructs are derived from one of the classes: \textit{people}, \textit{organizations}, \textit{technologies}, \textit{activities}, or \textit{software systems} \cite{sjoberg_building_2008}. We will use open coding \cite{stol_grounded_2016} to extract data.}
  \\  
  \cline{1-1}\cline{2-2}  
    \multicolumn{1}{|p{3cm}|}{Research Type (RQ1c)} &
    \multicolumn{1}{p{7cm}|}{For research type facets we use the taxonomy proposed by Wieringa \etal \cite{wieringa_requirements_2005}, containing the following categories: \textit{evaluation research}, \textit{solution proposal}, \textit{validation research}, \textit{philosophical paper}, \textit{opinion paper}, or \textit{experience paper}. Petersen \etal\cite{petersen_guidelines_2015} guidelines are used to assist in this categorization.}
  \\  
  \cline{1-1}\cline{2-2}  
    \multicolumn{1}{|p{3cm}|}{Empirical Evaluation (RQ1d)} &
    \multicolumn{1}{p{7cm}|}{Classification of the empirical study in the following categories of Wohlin \etal\cite{wohlin_experimentation_2012}: \textit{experiment/quasi-experiment}, \textit{case study}, or \textit{survey}.}
  \\ 
  \cline{1-1}\cline{2-2}  
    \multicolumn{1}{|p{3cm}|}{Limitations (RQ1e)} &
    \multicolumn{1}{p{7cm}|}{Limitations on the use of psychometric instruments (if exists). Such as what were the difficulties of adoption/application and data interpretation. We will use open coding \cite{stol_grounded_2016} to extract data.}
  \\ 
  \hline
 \end{tabular} }
\end{table}
\FloatBarrier

\subsection{Preliminary Results and Next Steps}
\label{sec:smp-progress}

\begin{figure}[h]
\caption{Current steps of this study}
\centering
\includegraphics[width=13cm]{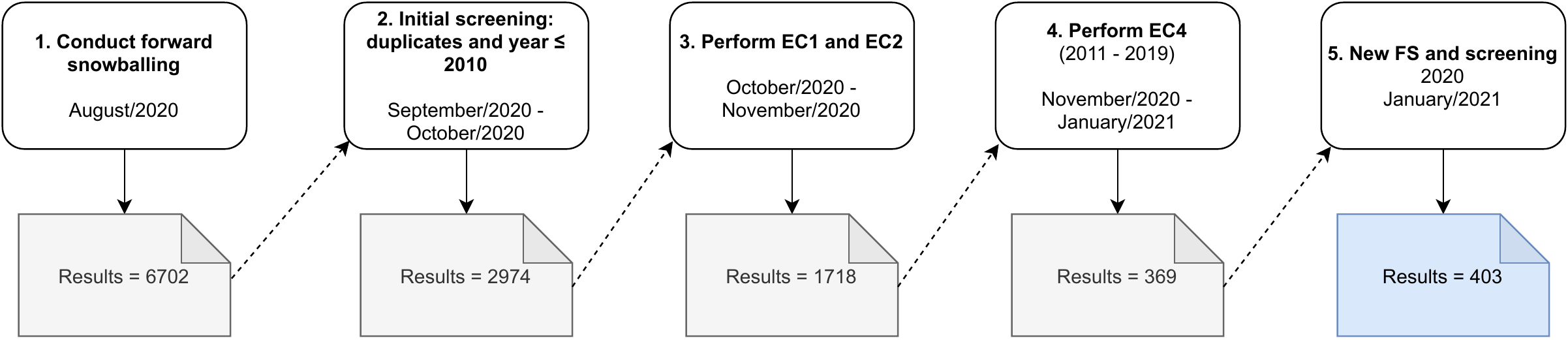}
\label{fig:current-step}
\end{figure}

The first step was to conduct FS in the seed set as described in section \ref{sec:smp-update}, which generated an entire of 6702 entries (step 1 of Figure \ref{fig:current-step}). Between September and October of 2020, the first author conducted an initial screening of duplicate and year less than or equal to 2010, given that Cruz \etal\cite{cruz_forty_2015} cover a range from 1970 to 2010.

Many entries were provided by Google Scholar/Publish or Perish 7 export feature with incomplete or incorrect data (\eg journal studies categorized in the entry as books or miscellaneous, or truncated title or abstracts). After removing duplicate entries and when possible, the first author manually collected the data for publication study and registered it in JabRef to perform a more reliable exclusion per year. The result of this initial screening resulted in 2974 entries (step 2 of Figure \ref{fig:current-step}).

Thereafter, the first author conducted another screening regarding the exclusion criteria EC1 and EC2. The removal was performed based on the metadata provided in the title, abstract, and journal/booktitle field entries. When it was not possible to easily identify, a verification was made through the URL of the entry or searching the source on the web. This exclusion was conducted between October and November of 2020 (step 3 of Figure \ref{fig:current-step}). Each entry was analyzed individually; the exclusion steps mentioned here reduced the set by 1718 entries.

To conduct the EC4 completely, we deleted the existing 2020 entries (which resulted in 369 entries of 2011 to 2019, step 4 of Figure \ref{fig:current-step}) and replaced them with the new 2020 FS ones to get a full ten-year index coverage. We applied the same previous ECs in this step, which resulted in a candidate set of 403 entries (step 5 of Figure \ref{fig:current-step}).

In the following we present some preliminary results extracted and organized by our RQs. So far, we mapped the years from 2011 to 2013\footnote{We refer to the identification of studies from our protocol as S(NUMBER), where (NUMBER) begins to account from 91 given that Cruz \etal\cite{cruz_forty_2015} has 90 studies.}. All the steps mentioned above involved the guidance and agreement of the second author. References of the selected studies are presented in Annex \ref{appendix:references}.

\subsubsection{Answers to the RQs}

\noindent\textbf{RQ1: }\rqmap

\begin{figure}[h]
\caption{Frequencies of psychometric instruments}
\centering
\includegraphics[width=12cm]{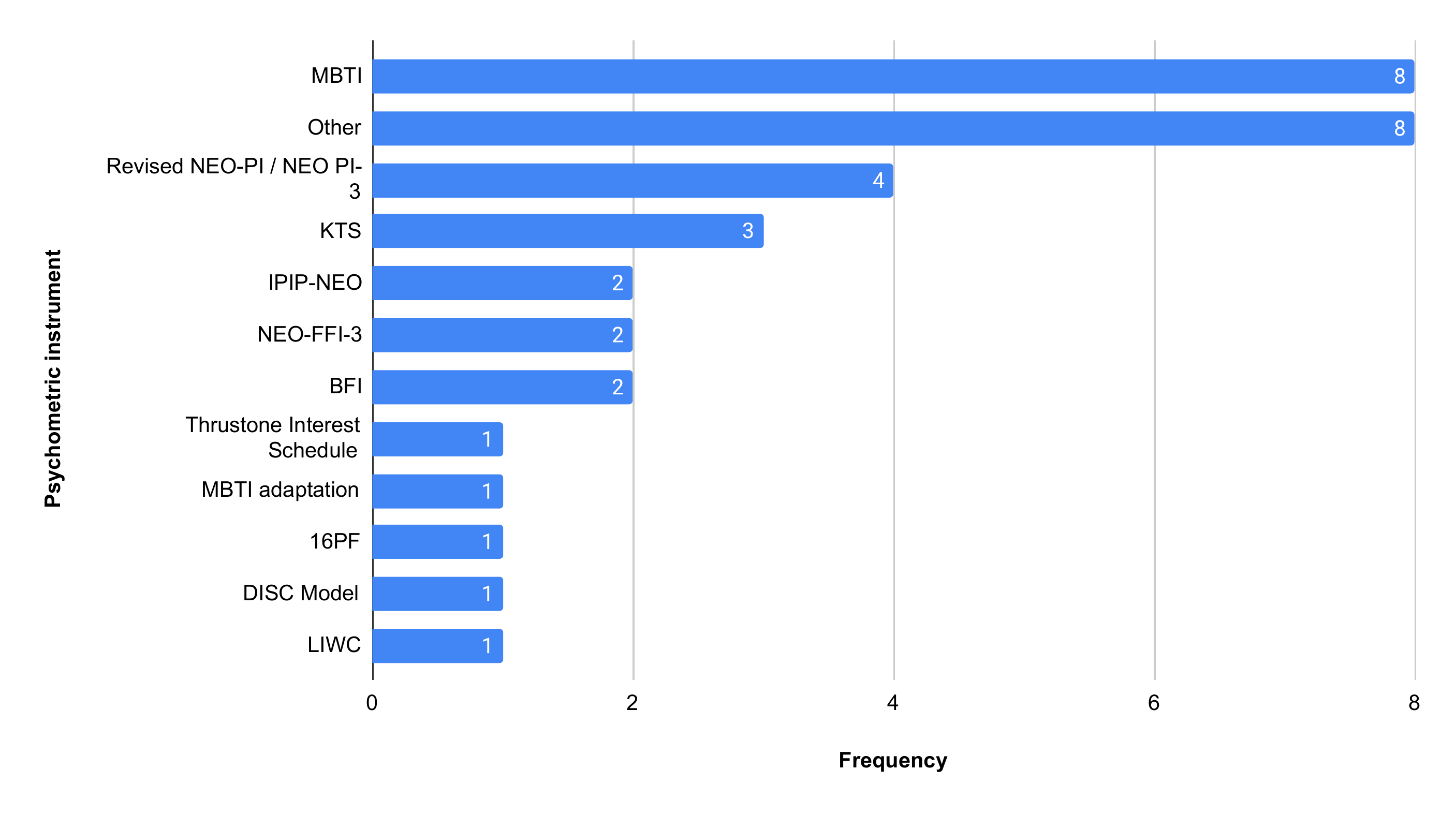}
\label{fig:pi-frequencies}
\end{figure}

We have identified 34 studies that employ psychometric instruments or personality theories, ranging from 2011 to 2013. The frequency of instruments is illustrated in Figure \ref{fig:pi-frequencies}, the most used ones were different versions of the Myers-Briggs Type Indicator (MBTI) [S91, S92, S96, S97, S98, S107, S110, S120], as can also be seen in Cruz \etal\cite{cruz_forty_2015}.

In sequence \textit{Other}, theories of personality such as the Five-Factor Model (also know as Big Five) are employed: [S94] uses it as a basis for the coding process in questionnaire responses; [S104] maps relationships between theory dimensions with soft and hard skills required from software engineers; [S116] is similar to the previous one, but the study uses subjects' social media posts metadata as a source. \textit{Other} was also used to categorize an instrument for which it was not possible to identify a version, bibliographic references, or obtain access [S102, S106, S109, S117] or that discusses the misuse in using psychometric instruments [S119].

To provide an overview of the guidelines raised by McDonalds and Edwards \cite{mcdonald_who_2007}, we have also extracted information from the details of types of instruments used (actual name, version, and theory) and how they were employed (from administration to data interpretation). Most studies did not report any bibliographic references and explicit versions of the instrument.

Furthermore, most studies also did not report anything different from ``we use x to measure personality'' regarding instruments application. Unlike them, [S93] mentions a consent form to participate in the experiment; [S95, S101] report that participants were given instructions; [S114, S118] report that interpretation procedures followed recommendations of the instruments' version. Beyond, [S114] claims that the results were given to the participants. [S124] reports that interviews were conducted following the administration to obtain results but without qualification details of the interviewer.
\\

\noindent\textbf{RQ1a: }\rqmapA

During the data extraction, it was possible to observe three major objectives.

\textit{Characterize software engineer personality:} some studies aimed to discover the personality of the software engineer [S91, S96, S97], and to compare them with the personality of professionals from other disciplines [S94, S95]; to discover personality patterns through a methodology [S98, S99]; to associate personality traits to roles in the software development activities (\eg developer, requirements analyst, tester) or required skills [S99, S104, S107, S116, S117, S123, S124]. Cultural differences were also investigated [S120].

\textit{Predicting team performance:} four studies used information on personality traits (along with other information or not) to predict a development team's performance to optimize resource use. One case reports for academic purposes, combining academic info to build teams to carry out a project [S92]. Others propose tools and methods using machine learning [S103], computational intelligence [S105], and rule-based approaches [S106].

\textit{Investigate the effect of personality:} data on personality traits are used as an intervention to investigate phenomena. We identified this objective with several specificities. Studies, such as [S93, S101, S112, S115], investigate the performance of compositions of pair programming teams; [S100] the influence on the quality of software developed and satisfaction of work perceived by a team; [S102] the programmer's error-proneness while developing; [S108, S110] the use of support tools; [S111] academic team achievements; [S113, S122] predict preferences on development method or mode of working (either face-to-face or virtual); [S114] the performance of software testing; [S118] team members on success in a software project; [S121] implementing new technology. Also, there are discussions concerning future Automated Personality Classification research [S109].
\\

\noindent\textbf{RQ1b: }\rqmapB

In Tables \ref{tab:ac-actor} to \ref{tab:ac-softwaresystem} we summarize the constructs to which the psychometric instruments are related by archetypal classes. We highlight the use of the framework to describe theories \cite{sjoberg_building_2008} and its archetypal classes as support for open coding \cite{stol_grounded_2016} constructs: an \textit{actor} applies \textit{technology} (we believe that \textit{intervention} is more appropriate in the context of psychometric instruments) to perform certain \textit{activities} on a \textit{software system}.

\begin{figure}[h]
\caption{SE constructs related to the psychometric instruments}
\centering
\includegraphics[width=13cm]{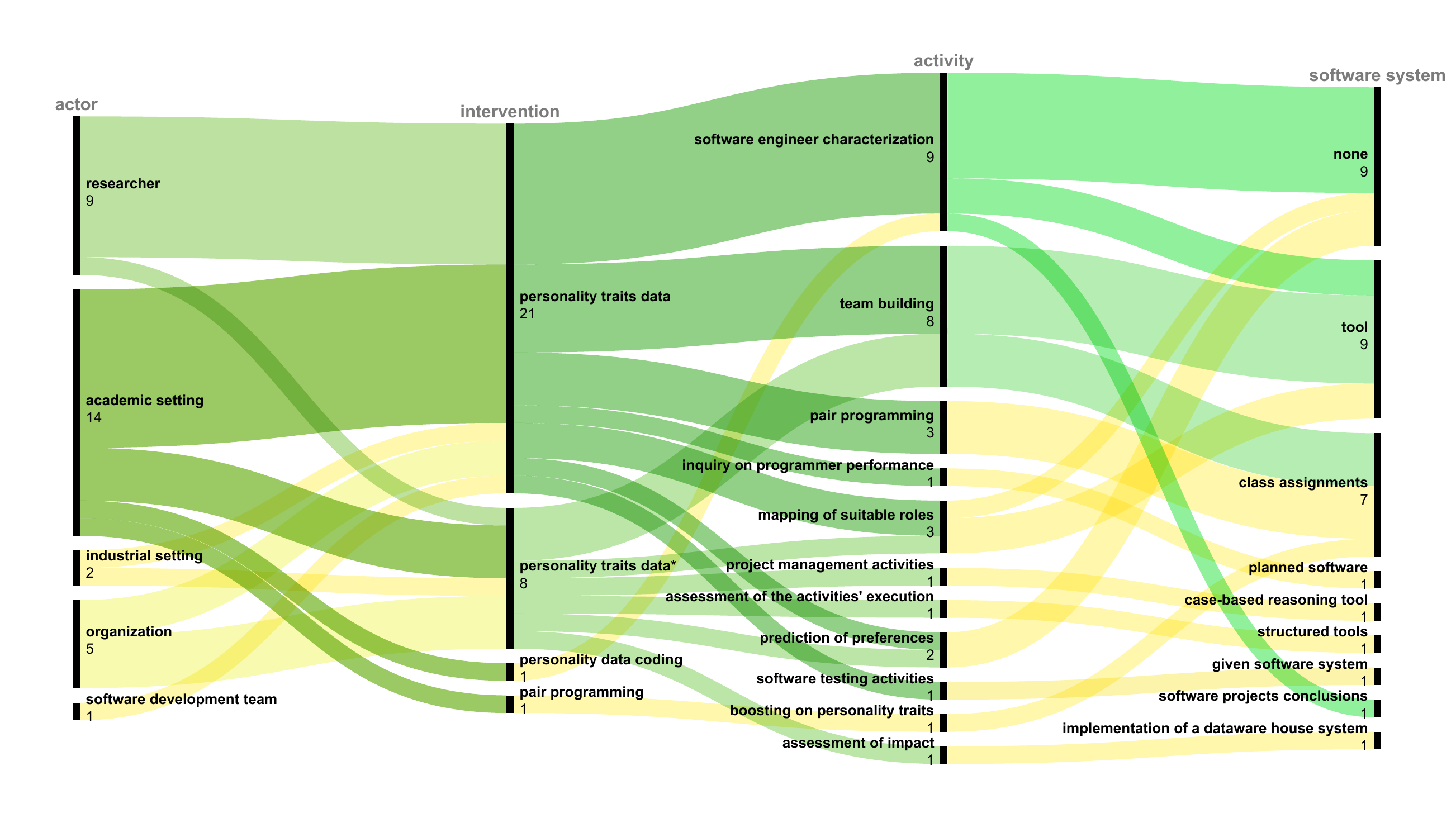}
\label{fig:pi-constructs}
\end{figure}

Figure \ref{fig:pi-constructs} depicts the relationship network between the constructs described in Tables \ref{tab:ac-actor} to \ref{tab:ac-softwaresystem}. It is possible to observe that constructs of class \textbf{actor} \textit{researcher} and \textit{academic setting} are the majority, indicating the application of studies in an academic setting or for purely investigative purposes by researchers, which corroborates the classification seen in RQ1c and RQ1d. In the \textbf{intervention class}, most data from personality traits measured by some psychometric instrument are used for investigation, followed by its increase with other data (such as leadership style, cognitive style, and skills required) indicated by the construct \textit{personality traits data*}. Regarding the \textbf{activity} class, the constructs \textit{software engineering characterization} and \textit{team building} are largely related to \textbf{interventions} of \textit{personality traits data}, holding the highest frequency of studies.

Still, these \textbf{activities} described earlier are also strongly related to the constructs of the \textbf{software system} class. It is possible to observe that \textit{software engineer characterization} is typically not related to any specific software system (\ie \textit{None} in Figure \ref{fig:pi-constructs}), indicating no direct reference to software system in these studies. The software system code \textit{Tool} indicates the use of some technique using software/games/logic rules to supported the studies. Moreover, \textit{class assignments} were specially related to \textit{team building} and \textit{team building}, where teams were built based in academic contexts based on personality data.

Please note that RQ1b aims at answering what parts of SE theory the psychometric instruments are related to. There may be similarities with the overall objectives of the conducted studies (RQ1a), but RQ1b is specifically focused on the instruments and their relations to SE theory elements.

\begin{table}[h]
\footnotesize
\centering
\caption{Actor}
\label{tab:ac-actor}
\begin{tabular}{|p{1.75cm}|p{8.25cm}|p{0.9cm}|}
\hline
\textbf{Class}\centering & \textbf{Description}\centering & \textbf{Count} \\ \hline
Researcher & The study author(s) act primarily for investigative purposes {[}S91, S96, S97, S104, S113, S117, S118, S120, S123{]}. & 9 \\ \hline
Academic setting & The study has an educational purpose or is applied in an academic setting due to scope limitations {[}S92, S93, S94, S100, S101, S107, S110, S111, S112, S114, S115, S116, S122, S124{]}. & 14 \\ \hline
Industrial setting & The study is applied in a industrial setting {[}S95, S121{]}. & 2 \\ \hline
Organization & The scope of the study is not clear (where does the study data come from?) {[}S98, S102, S103, S105, S108{]}. The code was adopted to be comprehensive. & 5 \\ \hline
Software development team (SDT) & An SDT was the interventor in the study {[}S99{]}. & 1 \\ \hline
\end{tabular}
\end{table}
\FloatBarrier

\begin{table}[]
\footnotesize
\centering
\caption{Intervention}
\label{tab:ac-intervention}
\begin{tabular}{|p{3cm}|p{7cm}|p{0.9cm}|}
\hline
\textbf{Class}\centering & \textbf{Description}\centering & \textbf{Count} \\ \hline
Personality traits data & The study has data on personality traits measured by some psychometric instrument {[}S91, S93, S95, S96, S97, S98, S99, S100, S101, S102, S107, S111, S112, S113, S114, S117, S118, S120, S123, S124{]} or other source {[}S116{]}. & 21 \\ \hline
Personality traits, academic, and software development methodology data & The study has as main input data of personality traits measured by some psychometric instrument, among other data {[}S92{]}. & 1 \\ \hline
Personality data coding & Personality data were used by means of coding based on some personality theory {[}S94{]}. & 1 \\ \hline
Personality traits and skills data & The study has as main input data of personality traits measured by some psychometric instrument and skills required for some software engineer role {[}S103, S104{]}. & 2 \\ \hline
Personality traits and nature of tasks data & The study has as main input data of personality traits measured by some instrument and already mapped information of personality traits recommended for execute a task in a software development process {[}S105{]}. & 1 \\ \hline
Personality traits and comprehension performance data & The study has as main input data of personality traits measured by some psychometric instrument and comprehension performance data {[}S110{]} & 1 \\ \hline
Personality traits and leadership style data & The study has as main input data of personality traits measured by some psychometric instrument, among other data {[}S121{]} & 1 \\ \hline
Personality and cognitive traits data & The study has as main input data of personality and cognitive traits measured by some psychometric instrument {[}S108, S122{]}. & 2 \\ \hline
Pair programming & The study adopted pair programming to assess some impact {[}S115{]}. & 1 \\ \hline
\end{tabular}
\end{table}
\FloatBarrier

\begin{table}[]
\footnotesize
\centering
\caption{Activities}
\label{tab:ac-activities}
\begin{tabular}{|p{2.5cm}|p{7.5cm}|p{0.9cm}|}
\hline
\textbf{Class}\centering & \textbf{Description}\centering & \textbf{Count} \\ \hline
Software engineer characterization & The study characterizes, in some extent, the software engineering professional. It means: comparison of SE and other professionals {[}S94, S95{]}, and discover (and/or compare) personalities {[}S91, S96, S97, S98, S118, S120, S123{]}. & 9 \\ \hline
Team building & The main activity of the study is to build software development teams with more than two members {[}S92, S99, S100, S103, S105, S107, S111, S124{]}. & 8 \\ \hline
Pair programming & The main activity of the study is to build pair programming teams {[}S93, S101, S112{]}. & 3 \\ \hline
Inquiry on programmer performance & The main activity of the study is to inquiry programmer performance {[}S102{]}. & 1 \\ \hline
Mapping of suitable roles & The main activity of the study is to mapping software engineer roles (developer, tester, project manager, etc.) {[}S104, S116, S117{]}. & 3 \\ \hline
Project management activities & The main activity of the study is to assess project manager performance given a software system {[}S108{]}. & 1 \\ \hline
Assessment of the activities' execution & The main activity of the study is to assess software artifacts and process in a software activity {[}S110{]}. & 1 \\ \hline
Prediction of preferences & The main activity of the study is to predict some preference of software engineer professional in a software development activity {[}S113, S122{]}. & 2 \\ \hline
Software testing activities & The main activity of the study is to assess tester performance given a software [S114]. & 1 \\ \hline
Boosting on personality traits & The main activity of the study is to boost personality traits at some level {[}S115{]}. & 1 \\ \hline
Assessment of impact & The main activity of the study is to assess impacts {[}S121{]}. & 1 \\ \hline
\end{tabular}
\end{table}
\FloatBarrier

\begin{table}[]
\footnotesize
\centering
\caption{Software system}
\label{tab:ac-softwaresystem}
\begin{tabular}{|p{2.5cm}|p{7.5cm}|p{0.9cm}|}
\hline
\textbf{Class}\centering & \textbf{Description}\centering & \textbf{Count} \\ \hline
None & No software system was used in the study {[}S91, S94, S95, S96, S97, S104, S113, S120, S122{]}. & 9 \\ \hline
Tool & The study used a machine learning/computational intelligence {[}S92, S98, S103, S105{]} some kind of algorithm {[}S116, S117, S123, S124{]}, or a gamecard {[}S99{]}. & 9 \\ \hline
Class assignments & The study used a software system for academic settings, previously developed for some specific purpose (be tested, refactored, ...) or developed during the conduction of the study by students {[}S93, S100, S101, S107, S111, S112, S115{]}. & 7 \\ \hline
Planned software & The study uses software (or requirements of it) that has not yet been developed, but it was during the conduction of the study {[}S102{]}. & 1 \\ \hline
Case-based reasoning (CBR) tool & The study used a CBR tool to perform an investigation {[}S108{]}. & 1 \\ \hline
Structured tools & The study used structured tools to perform an investigation {[}S110{]}. & 1 \\ \hline
Given software system & A specific software system was used in the study because it has some kind of problem to be investigated by the subjects. For instance: in software testing {[}S114{]}. & 1 \\ \hline
Software projects conclusion & The study used data from conclusions of software system projects (successfully or not) {[}S118{]}. & 1 \\ \hline
Implementation of a dataware house system & The study used data from implementation of a dataware house system {[}S121{]}. & 1 \\ \hline
\end{tabular}
\end{table}
\FloatBarrier

\noindent\textbf{RQ1c: }\textit{Which type of research do the studies refer for?}

Figure \ref{fig:pi-frequencies} shows the distribution of research types facets. It is possible to observe that \textit{validation research} (19 out of 32) overrepresented the set of mapped studies. This facet includes empirical studies, as the less frequent \textit{evaluation research} facet (6 out of 32). The difference indicates that most empirical research has been conducted in academic scenarios for initial validation purposes and does not evaluate something in industrial scenarios. Some \textit{solution proposals} have also been mapped (7 out of 32), which typically represent new proposals with some limited evaluation required for publication.

\begin{figure}[h]
\caption{Frequency of research type facets of Wieringa \etal\cite{wieringa_requirements_2005}}
\centering
\includegraphics[width=10cm]{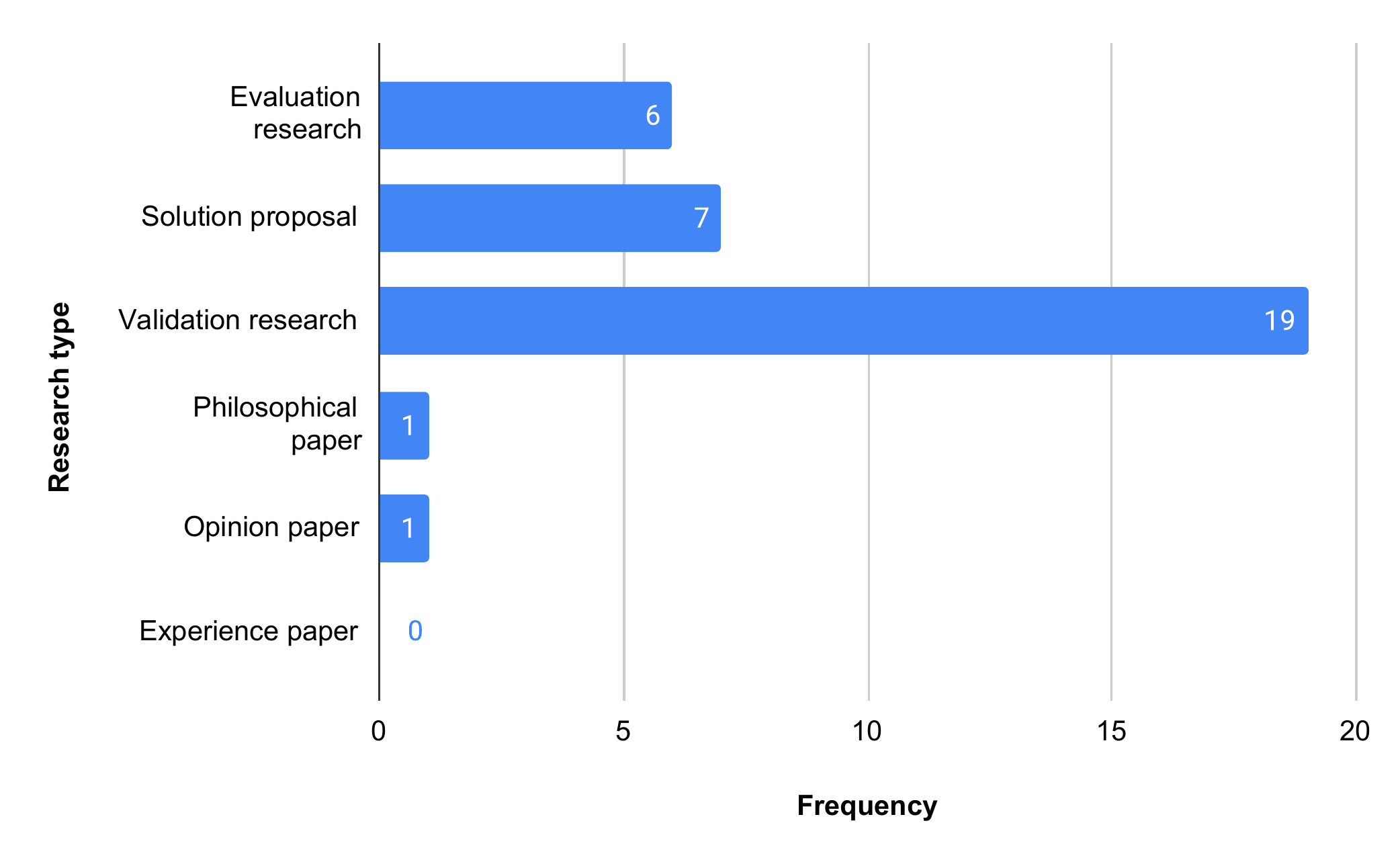}
\label{fig:rt-frequencies}
\end{figure}
\FloatBarrier

\noindent\textbf{RQ1d: }\rqmapD

\begin{figure}[h]
\caption{Frequency of empirical evaluations adopted \cite{wohlin_experimentation_2012}}
\centering
\includegraphics[width=10cm]{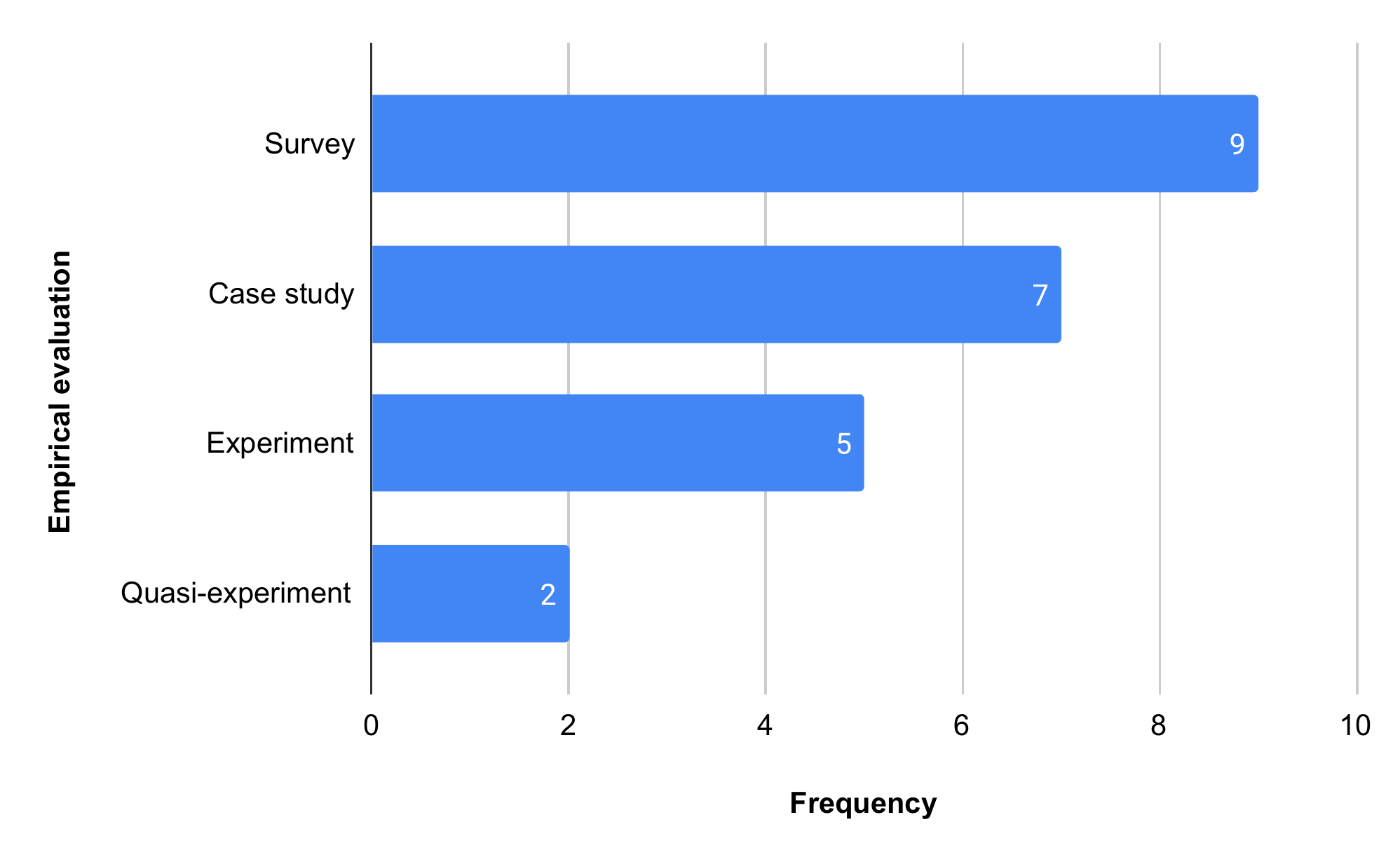}
\label{fig:ee-frequencies}
\end{figure}

Figure \ref{fig:ee-frequencies} depicts the frequency of empirical evaluations adopted (23 out of 32 studies). It is possible to observe that most of the studies (9 out of 23) used surveys, mostly to apply the psychometric instruments. Still, it is not possible to draw consistent conclusions, such as trends and gaps in empirical evaluation adoption given that we only mapped three years so far.

\FloatBarrier

\noindent\textbf{RQ1e: }\rqmapE

Some studies (16 out of 32) reported limitations related to adoption of psychometric instruments. An overview of these limitations is described below.

\textit{Possible misuse of psychometric instrument:} the authors of [S100] declare that adopting the psychometric instrument to build teams can affect the construct validity, but this limitation is mitigated by relying on the literature. Another study that we consider noteworthy to mention reports on the wide and mistaken use of MBTI for professional guidance in careers in computing [S119]. The authors report a discussion with experts that it could be an ``abuse of the MBTI licensing requirements'' and ineffective for the purpose.

\textit{Bias in subject responses:} the authors indicate that subjects' self-administration of psychometric instruments can become a threat if there is no honesty in their responses [S101, S112, S122]. This limitation is mitigated in these studies by ensuring they made aware that the response data obtained is anonymized and used only for research purposes.

\textit{Statistical power of psychometric instrument:} Regarding this limitation, the increase of personality traits data by adopting more psychometric instruments for better prediction is reported [S98, S124]. Other authors also report on replacing the instrument in future work, given that it was not possible to draw significant conclusions in the use of a specific instrument [S113].

Still, regarding the statistical structure of the instrument, [S108] uses an instrument based on the Five-Factor Model and recommends in future works to use the MBTI. In turn, [S112] uses an instrument based on Jung's theory (KTS / MBTI), which provides strict personality measurements in dichotomies (\eg introvert-extrovert). This leads to more than one person being categorized in one dichotomy even if they hold different levels. Hereafter, we can notice inconsistencies in the literature regarding the statistical power of the instruments reported in the limitations of the studies.

\textit{No representative data input/no representative sample:} Restrictions of representative data to predict team performance are reported [S92, S105]; the sample size was sufficient to control more than one personality dimension in the experiment, only the Openness to Experience dimension of the Five-Factor Model was considered [S93]; in [S111, S113], a larger sample could add more statistical power to provide other conclusions; limitations in the empirical method employed did not allow representative statistical analysis to generalize results [S118, S122].

\textit{Paid subjects:} Participants were paid to participate in the study, which may have influenced them somehow [S114].

\section{Preliminary Survey Protocol}
\label{sec:psp}

This section presents our preliminary survey protocol (primary study) that we aim to conduct to achieve the main objective of this study described in the introduction. We adopted guidelines for conducting surveys in SE \cite{linaker_guidelines_2015} and advice to overcome common challenges \cite{wagner_challenges_2020}, to assist in developing this protocol. This protocol was developed by the first author and reviewed by the second.

The survey's main goal is to collect the point of view of social science researchers on the use of psychometric instruments in SE, as mapped through our ongoing secondary study described in section \ref{sec:smp}. We adopted the Goal-Question-Metric (GQM) method \cite{wohlin_experimentation_2012} to characterize the goal in Table \ref{table:GQM-survey}:

\begin{table}[h] \footnotesize
\centering
\caption{Goal of the survey}
\resizebox{\textwidth}{!}{%
\label{table:GQM-survey}
\begin{tabular}{ll}
\hline
 \textbf{Analyze} & the
adoption of psychometric 
instruments regarding personality in SE\\ 
 \textbf{For the purpose of} & characterization\\
 \textbf{With respect to} & their correct use \\
 \textbf{From the point of view of the} & social
sciences researchers  \\
 \textbf{In the context of}  & the mapped
SE research  \\ \hline
\end{tabular}%
}
\end{table}

Hence, given that there is a poor collaboration between SE and social sciences researchers, and misuse of psychometric instruments, as pointed out in sections \ref{sec:introduction}, \ref{sec:background-bse}, and \ref{sec:background-psySE}, we have elaborated the following RQ in order to address the survey's main goal:

\begin{center}
\textbf{RQ2: }\rqsurvey
\end{center}

The remainder of this section is organized as follows. Subsection \ref{sec:survey-population} presents information on population and sample. Subsection \ref{sec:survey-instrument} presents the preliminary survey instrument design. Subsections \ref{sec:survey-dataanalisys} and \ref{sec:survey-currents} discuss data analysis and current status, respectively.

\subsection{Population and Sample}
\label{sec:survey-population}

The population selected for the study are social science researchers with an area of interest in psychometrics. Initially, we are planning to collaborate with social science researchers from the Department of Psychology at the Pontifical Catholic University of Rio de Janeiro (PUC-Rio)\footnote{http://www.psi.puc-rio.br/} to help us to best plan the outline the population and identify a representative sample. 

A priori, the sample may be characterized as being non-probabilistic and accidental, as frequent in SE, where researchers recruit subjects based on personal connections \cite{linaker_guidelines_2015}. To characterize a sample frame as ideal, we use points highlighted by Linåker \etal\cite{linaker_guidelines_2015}.

\begin{itemize}
    \item \textbf{Target audience:} social science researchers;
    \item \textbf{Unit of analysis:} psychometric instruments adopted in SE research regarding personality;
    \item \textbf{Unit of observation:} social sciences researchers with an area of interest in psychometrics;
    \item \textbf{Search unit:} to be defined;
    \item \textbf{Source of sampling:} to be defined.
\end{itemize}

\subsection{Instrument Design}
\label{sec:survey-instrument}

\begin{figure}[h]
\caption{Illustration of clustering of the systematic mapping study artifacts}
\centering
\includegraphics[width=10cm]{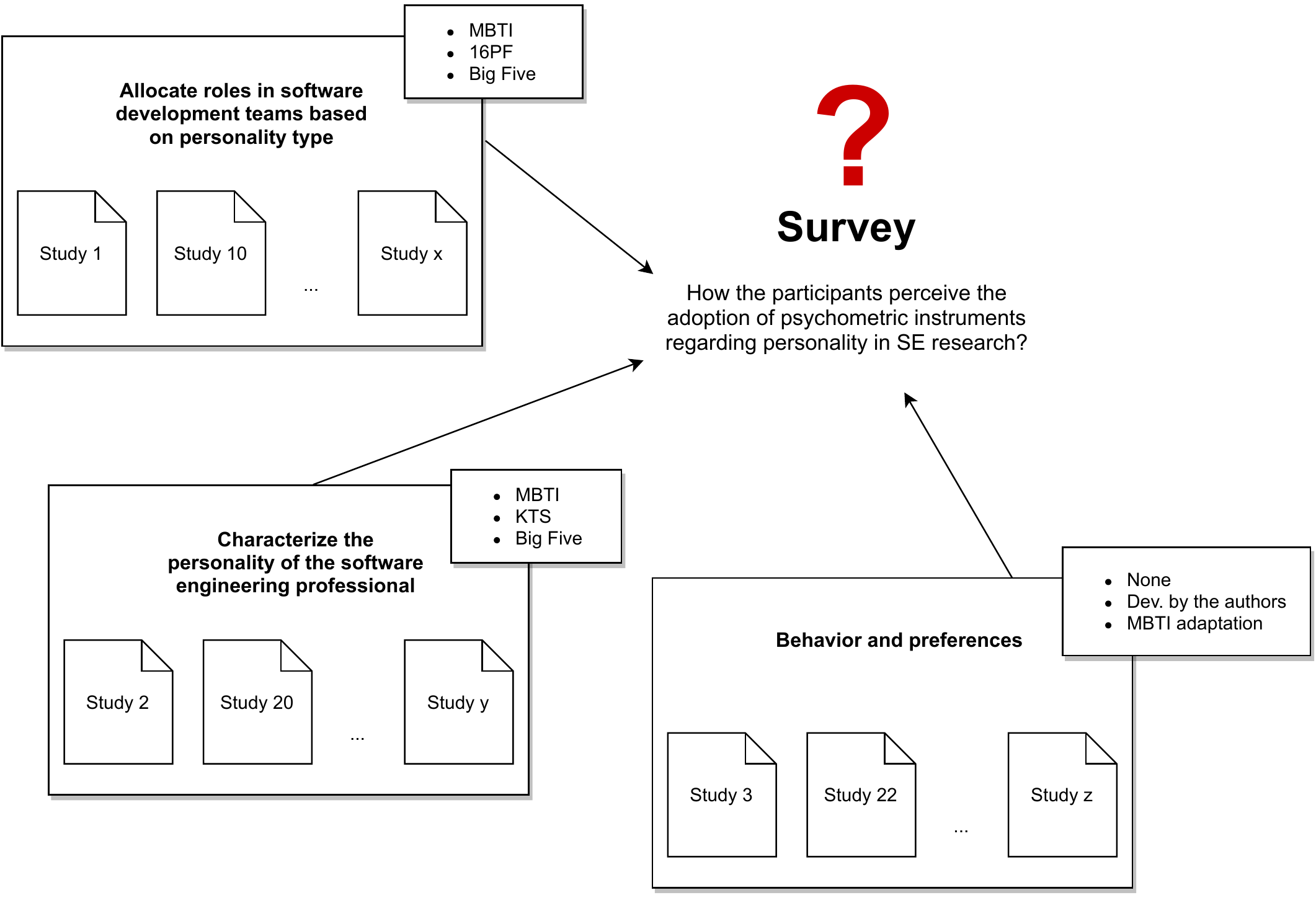}
\label{fig:survey-strategy}
\end{figure}

The input to conduct the survey consists of completing the systematic mapping. Thus, to assist in surveying the subjects, we intend to cluster the final set of primary studies by the psychometric instrument's purposes identified through the answered systematic mapping \textbf{RQ1}. Figure \ref{fig:survey-strategy} illustrates this: with each identified instrument (MBTI, 16PF, and a Big Five based instrument), and a set of identified purposes (\eg \textit{allocate roles in software development teams based on personality type}), and the list of bibliographic references (if the subject needs to consult details of a specific study).

We rely on the Technology Acceptance Model (TAM) \cite{davis_perceived_1989} to assess the psychometric instruments and their use in SE research, as we see them as technologies/tools. Thus, the survey questionnaire presented in Table \ref{table:survey-questions} aims to capture the \textit{perception of their usefulness} (questions $Q5$ and $Q6$), and \textit{perception of ease of use} (questions $Q7$ and $Q8$). Questions $Q1$-$Q4$ concern demographic information on the social sciences researcher (subject) that we intend to collect. Finally, questions $Q9$ and $Q10$ capture additional limitations regarding the application process and recommendations. 

\begin{table}[] \footnotesize
\centering
\caption{Survey questionnaire}
\label{table:survey-questions}
\begin{tabular}{|l|l|}
\hline
\textit{\textbf{ID}} & \textit{\textbf{Questions}} \\ \hline
\multicolumn{2}{|l|}{Demographic questions} \\ \hline
Q1 & \begin{tabular}[c]{@{}l@{}}What is your current academic degree in social sciences (B.Sc., M.Sc., Ph.D.)? \end{tabular} \\ \hline
Q2 & \begin{tabular}[c]{@{}l@{}}How long (in years) do you hold a social sciences bachelor's degree (such as in psychology)? \end{tabular} \\ \hline
Q3 & \begin{tabular}[c]{@{}l@{}}How long (in years) have you been researching in the field of psychometrics? \end{tabular} \\ \hline
Q4 & \begin{tabular}[c]{@{}l@{}}How long (in years) have you employed psychometric instruments in practice? \end{tabular} \\ \hline

\multicolumn{2}{|l|}{\begin{tabular}[c]{@{}l@{}}RQ2: How the participants perceive the adoption of psychometric instruments regarding per-\\sonality in SE research?\\ Given the set of psychometric instruments {[}$x_1, x_2, ..., x_n${]} used in purpose in SE research\\ {[}$y_1, y_2, ..., y_n${]}, please answer:\end{tabular}} \\ \hline

Q5 & \begin{tabular}[c]{@{}l@{}}Using the instrument $X$ with the purpose $Y$ in SE research is consistent with its\\ original proposal in social sciences. \\ (strongly agree, agree, neutral, disagree, strongly disagree). \\ Please detail your answer:\end{tabular} \\ \hline
Q6 & \begin{tabular}[c]{@{}l@{}}Using the  instrument $X$ can support the SE researcher to achieve the purpose $Y$.\\ (strongly agree, agree, neutral, disagree, strongly disagree) \\ Please detail your answer:\end{tabular} \\ \hline
Q7 & \begin{tabular}[c]{@{}l@{}}Instrument $X$ is easy to apply to achieve the purpose $Y$ \\ (strongly agree, agree, neutral, disagree, strongly disagree)\\ Please detail your answer:\end{tabular} \\ \hline
Q8 & \begin{tabular}[c]{@{}l@{}}Analyzing and interpreting the result of using instrument $X$ with the purpose $Y$ \\ is easy for a non-social science researchers. \\ (strongly agree, agree, neutral, disagree, strongly disagree) \\ Please detail your answer:\end{tabular} \\ \hline
Q9 & \begin{tabular}[c]{@{}l@{}}The application process of the psychometric \\ instrument $X$ with the purpose $Y$ is consistent\\ (strongly agree, agree, neutral, disagree, strongly disagree).\\ Please detail your answer:\end{tabular} \\ \hline
Q10 & Are there any recommendations to improve the negative points? \\ \hline
\end{tabular}
\end{table}

\subsection{Data Analysis}
\label{sec:survey-dataanalisys}

As presented in Table~\ref{table:survey-questions}, in our preliminary questionnaire, we only deal with rational data with regard to the experience in years. For such data, we intend to use the mean and standard deviation. All the remaining questions concern ordinal scales (\eg academic degree, or Likert scales) or open-ended questions. We intend to analyze the agreement of the subjects, in the ordinal data, by counting frequencies, using the median (central tendency) and mode (most frequent answer), potentially also using the median absolute deviation (depending on the sample size). We do not discard using Bootstrapping confidence intervals, but this will heavily depend on our sample size. Cohen's Kappa will be applied for inter-rater consistency evaluations. As for the data visualization, we will use histograms and pie charts to analyze the distribution of answers. For open-ended questions, we will use open (and potentially axial) coding procedures \cite{stol_grounded_2016} to extract relevant data.

\subsection{Current Status and Next Steps}
\label{sec:survey-currents}

\begin{figure}[h]
\caption{Next steps to refine survey}
\centering
\includegraphics[width=10cm]{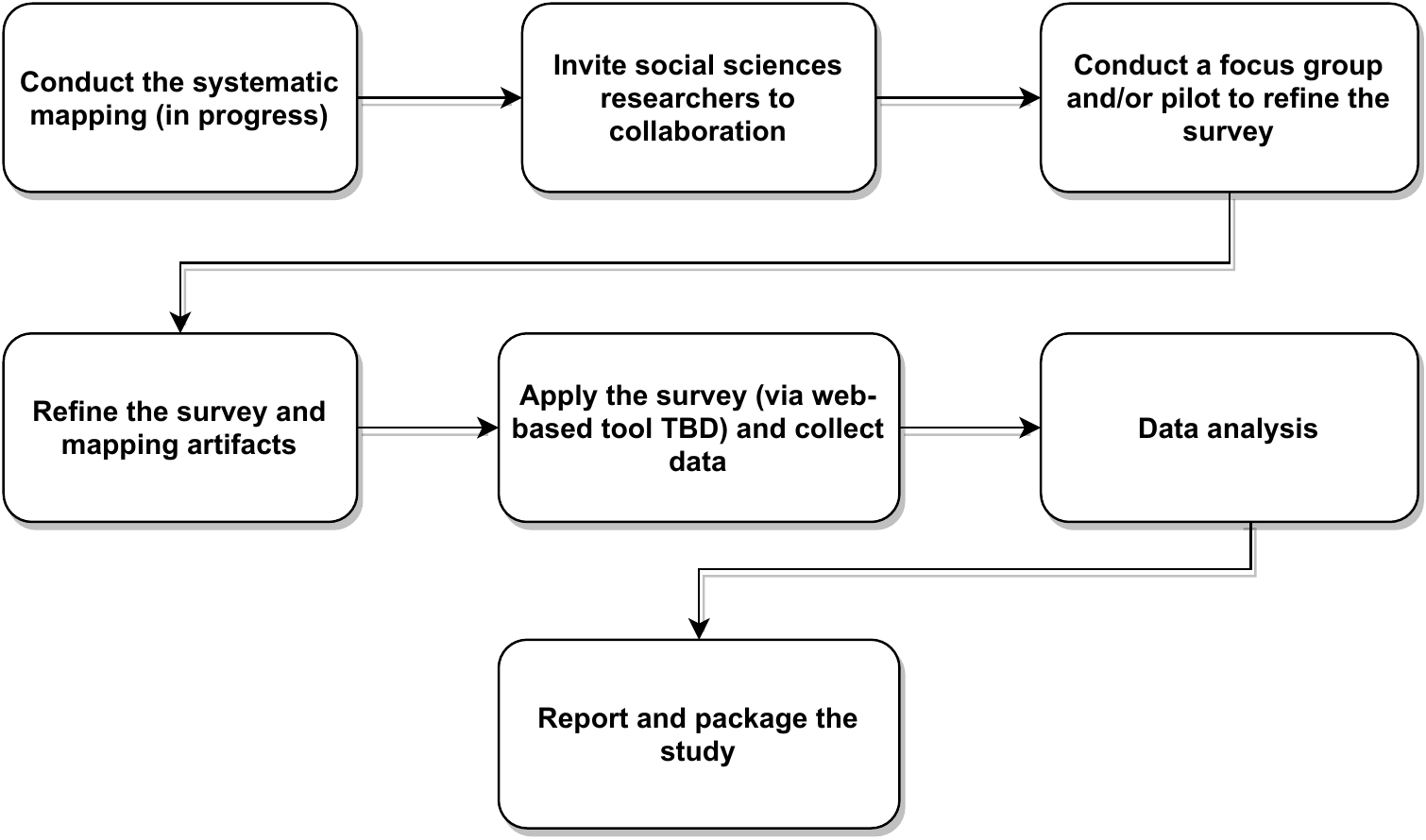}
\label{fig:survey-process}
\end{figure}

In this section we presented our preliminary survey protocol. Currently, we are conducting the data extraction from the candidate set (see Table \ref{IE-table}) in the systematic mapping. The next steps for this survey can be seen in Figure \ref{fig:survey-process}.

\begin{itemize}
\item \textbf{Invite social sciences researchers to collaboration}: we intend to discuss with social sciences researchers about: better definition of the sample; execution of a focus group and/or pilot and data collection agenda.
\item \textbf{Conduct a focus group and/or pilot study to refine the survey:} with a defined agenda, we plan to conduct a focus group and/or pilot process in a subgroup of respondents. We aim obtain feedback to make refinements on the survey instrument.
\item \textbf{Refine the survey and mapping artifacts:} with the previous step already executed and discussed, if possible, in collaboration with the social sciences researchers, we will redesign the survey instrument.
\item \textbf{Apply the survey and collect data:} with a schedule defined, we intend to apply the survey. 
\item \textbf{Data analysis:} the data collected from the survey will be analyzed with the support of the statistical techniques briefly discussed (open to changes).
\item \textbf{Report and package the study:} finally, we intend to report the results of our study and make the artifacts available in order to make them auditable and replicable.

\end{itemize}

\section{Conclusions and Next Steps of The Study}
\label{sec:timetable}

As presented in subsections \ref{sec:smp-progress} and \ref{sec:survey-currents}, this study is at an initial step. Thus, we cannot draw major discussions and conclusions yet. The next steps direct us to finish the data extraction for the next seven years (2014 - 2020) according to the mapping protocol's inclusion and exclusion criteria (present in Table \ref{IE-table}) and thus conduct data extraction as described in the Data Extraction Form (see Table \ref{DEF}). As soon as possible, and with a considerable amount\footnote{To be discussed among the authors.} of information extracted, we intend to refine the survey protocol with social sciences researchers as discussed in subsection \ref{sec:survey-currents}. Threats to the study's validity need to be strongly discussed among the authors and were not discussed in this ongoing study.

\clearpage

\addtocontents{toc}{\protect\setcounter{tocdepth}{1}}
\addcontentsline{toc}{section}{References}
\bibliography{references.bib}
\bibliographystyle{ieeetr}

\pagebreak
\begin{appendices}
\section{List with the final set of included studies (ongoing)}
\appendix
\label{appendix:references}

\noindent[S91] Varona, D., Capretz, L.F., Piñero, Y., 2011. Personality types of Cuban software developers. Global Journal of Engineering Education 13, 5.

\noindent[S92] Omar, M., Syed-Abdullah, S.-L., Mohd Hussin, N., 2011. Developing a Team Performance Prediction Model: A Rough Sets Approach, in: Abd Manaf, A., Zeki, A., Zamani, M., Chuprat, S., El-Qawasmeh, E. (Eds.), Informatics Engineering and Information Science, Communications in Computer and Information Science. Springer, Berlin, Heidelberg, pp. 691–705. https://doi.org/10.1007/978-3-642-25453-6\_58

\noindent[S93] Salleh, N., Mendes, E., Grundy, J., 2011. The effects of openness to experience on pair programming in a higher education context, in: 2011 24th IEEE-CS Conference on Software Engineering Education and Training (CSEE T). Presented at the 2011 24th IEEE-CS Conference on Software Engineering Education and Training (CSEE T), pp. 149–158. https://doi.org/10.1109/CSEET.2011.5876082

\noindent[S94] Alexander, P., Pieterse, V., Lotriet, H., 2011. A Comparison Of Computing And Non-computing Students’ Personalities Based On The Five-Factor Model, in: ECIS 2011 Proceedings. Presented at the European Conference on Information Systems (ECIS).

\noindent[S95] Monika Mahindra, K. BharathiKumari, 2011. Personality and interest pattern of men in different occupations at Escorts Limited, Faridabad. Global Journal of Research in Management 1.

\noindent[S96] Raza, A., UlMustafa, Z., Capretz, L., 2011. Personality Dimensions and Temperaments of Engineering Professors and Students – A Survey. Journal of Computing 3, 13–20.

\noindent[S97] Varona, D., Capretz, L.F., 2011. Comparing Cuban and Brazilian software engineers. World Transactions on Engineering and Technology Education 9, 5.

\noindent[S98] Martínez, L.G., Castro, J.R., Licea, G., Rodríguez-Díaz, A., Alvarez, C.F., 2021. Knowing Software Engineer’s Personality to Improve Software Development. Presented at the 6th International Conference on Software and Data Technologies, pp. 99–104.

\noindent[S99] Yilmaz, M., OConnor, R.V., 2012. Towards the Understanding and Classification of the Personality Traits of Software Development Practitioners: Situational Context Cards Approach, in: 2012 38th Euromicro Conference on Software Engineering and Advanced Applications. Presented at the 2012 38th Euromicro Conference on Software Engineering and Advanced Applications, pp. 400–405. https://doi.org/10.1109/SEAA.2012.62

\noindent[S100] Gómez, M.N., Acuña, S.T., Genero, M., Cruz-Lemus, J.A., 2012. How Does the Extraversion of Software Development Teams Influence Team Satisfaction and Software Quality?: A Controlled Experiment. International Journal of Human Capital and Information Technology Professionals (IJHCITP) 3, 11–24. https://doi.org/10.4018/jhcitp.2012100102

\noindent[S101] Sfetsos, P., Adamidis, P., Angelis, L., Stamelos, I., Deligiannis, I., 2012. Investigating the Impact of Personality and Temperament Traits on Pair Programming: A Controlled Experiment Replication, in: 2012 Eighth International Conference on the Quality of Information and Communications Technology. Presented at the 2012 Eighth International Conference on the Quality of Information and Communications Technology, pp. 57–65. https://doi.org/10.1109/QUATIC.2012.36

\noindent[S102] Amir Asil, Hasan Asil, Mohsen Asil, 2012. The Effect of Programmers on Software Project Management Based Upon Personality Five Traits Theory. IJETAE 2, 3.

\noindent[S103] Stylianou, C., Andreou, A.S., 2012. A Multi-objective Genetic Algorithm for Software Development Team Staffing Based on Personality Types, in: Iliadis, L., Maglogiannis, I., Papadopoulos, H. (Eds.), Artificial Intelligence Applications and Innovations, IFIP Advances in Information and Communication Technology. Springer, Berlin, Heidelberg, pp. 

\noindent[S104] Rehman, M., Mahmood, A.K., Salleh, R., Amin, A., 2012. Mapping job requirements of software engineers to Big Five Personality Traits, in: 2012 International Conference on Computer Information Science (ICCIS). Presented at the 2012 International Conference on Computer Information Science (ICCIS), pp. 1115–1122.

\noindent[S105] Stylianou, C., Gerasimou, S., Andreou, A.S., 2012. A Novel Prototype Tool for Intelligent Software Project Scheduling and Staffing Enhanced with Personality Factors, in: 2012 IEEE 24th International Conference on Tools with Artificial Intelligence. Presented at the 2012 IEEE 24th International Conference on Tools with Artificial Intelligence, pp. 277–284. https://doi.org/10.1109/ICTAI.2012.45

\noindent[S106] Omar, M., Syed-Abdullah, S.-L., Hussin, N.M., 2012. eTiPs: A Rule-based Team Performance Prediction Model Prototype. Procedia Technology, First World Conference on Innovation and Computer Sciences (INSODE 2011) 1, 390–394.

\noindent[S107] Montequín, V.R., Balsera, J.V., Fernández, J.M.M., Nieto, A.G., 2012. Using Myers-Briggs Type Indicator (MBTI) as a Tool for Setting up Student Teams for Information Technology Projects. Journal of Information Technology and Application in Education 7.

\noindent[S108] Mair, C., Martincova, M., Shepperd, M., 2012. An Empirical Study of Software Project Managers Using a Case-Based Reasoner, in: 2012 45th Hawaii International Conference on System Sciences. Presented at the 2012 45th Hawaii International Conference on System Sciences, pp. 1030–1039. https://doi.org/10.1109/HICSS.2012.96

\noindent[S109] Kartelj, A., Filipović, V., Milutinović, V., 2012. Novel approaches to automated personality classification: Ideas and their potentials, in: 2012 Proceedings of the 35th International Convention MIPRO. Presented at the 2012 Proceedings of the 35th International Convention MIPRO, pp. 1017–1022.

\noindent[S110] Gorla, N., Chiravuri, A., Meso, P., 2013. Effect of personality type on structured tool comprehension performance. Requirements Eng 18, 281–292.

\noindent[S111] Koroutchev, K., Acuña, S.T., Gómez, M.N., 2013. The Social Environment as a Determinant for the Impact of the Big Five Personality Factors and the Group’s Performance. International Journal of Human Capital and Information Technology Professionals (IJHCITP) 4, 1–8. https://doi.org/10.4018/jhcitp.2013010101

\noindent[S112] Panagiotis Sfetsos, Panagiotis Adamidis, Lefteris Angelis, Ioannis Stamelos, Ignatios Deligiannis, 2013. Heterogeneous Personalities Perform Better in Pair Programming: The Results of a Replication Study. SQP 15, 4–15.

\noindent[S113] Bishop, D., 2013. Personality Theory as a Predictor for Agile Preference, in: MWAIS 2013 Proceedings. Presented at the MWAIS conference.

\noindent[S114] Kanij, T., Merkel, R., Grundy, J., 2013. An empirical study of the effects of personality on software testing, in: 2013 26th International Conference on Software Engineering Education and Training (CSEE T). Presented at the 2013 26th International Conference on Software Engineering Education and Training (CSEE T), pp. 239–248.

\noindent[S115] Radhakrishnan, P., Kanmani, S., 2013. Improvement of programming skills using pair programming by boosting extraversion and openness to experience. IJTCS 4, 13.

\noindent[S116] Kasturi Dewi Varathan, Li Thing Thiam, 2013. Mining Facebook in Identifying Software Engineering Students’ Personality and Job Matching, in: ACSET2013 Conference Proceedings. Presented at the The Asian Conference on Society, Education and Technology, Osaka, Japan, pp. 278–279.

\noindent[S117] Abdul Rehman Gilal, Mazni Omar, Kamal Imran Sharif, 2013. Discovering Personality Types and Diversity Based on Software Team Roles, in: Proceedings of the 4th International Conference on Computing and Informatics. Presented at the International Conference on Computing and Informatics (ICOCI), Sarawak, Malaysia, Universiti Utara Malaysia, p. 6.

\noindent[S118] {Karapıçak, Ç.M., Demirörs, O.}, 2013. A Case Study on the Need to Consider Personality Types for Software Team Formation, in: Woronowicz, T., Rout, T., O’Connor, R.V., Dorling, A. (Eds.), Software Process Improvement and Capability Determination, Communications in Computer and Information Science. Springer, Berlin, Heidelberg, pp. 120–129. https://doi.org/10.1007/978-3-642-38833-0\_11

\noindent[S119] McEwan, T., McConnell, A., 2013. Young people’s perceptions of computing careers, in: 2013 IEEE Frontiers in Education Conference (FIE). Presented at the 2013 IEEE Frontiers in Education Conference (FIE), pp. 1597–1603.

\noindent[S120] Varona, D., Capretz, L.F., Raza, A., 2013. A multicultural comparison of software engineers. WIETE 11, 5.

\noindent[S121] Thiti Phiriyayotha, Siriluck Rotchanakitumnuai, 2013. Data Warehouse Implementation Success Factors and the Impact of Leadership and Personality on the Relationship between Success Factors. Journal of Business and Economics 4, 9.

\noindent[S122] Luse, A., McElroy, J.C., Townsend, A.M., DeMarie, S., 2013. Personality and cognitive style as predictors of preference for working in virtual teams. Computers in Human Behavior 29, 1825–1832. https://doi.org/10.1016/j.chb.2013.02.007

\noindent[S123] Bazelli, B., Hindle, A., Stroulia, E., 2013. On the Personality Traits of StackOverflow Users, in: 2013 IEEE International Conference on Software Maintenance. Presented at the 2013 IEEE International Conference on Software Maintenance, pp. 460–463.

\noindent[S124] Martínez, L.G., Licea, G., Rodríguez, A., Castro, J.R., Castillo, O., 2013. Using MatLab’s fuzzy logic toolbox to create an application for RAMSET in software engineering courses. Computer Applications in Engineering Education 21, 596–605.

\end{appendices}

\end{document}